\documentclass[onecolappendix]{aastex6}

\begin{document}
\title{Are there moonlets near the Uranian $\alpha$ and $\beta$ rings?}
\author{R. O. Chancia and M. M. Hedman}
\affil{\textit{Department of Physics, University of Idaho, Moscow ID 83844-0903}}
\email{rchancia@uidaho.edu}

\begin{abstract}
The \textit{Voyager 2} Radio Science Subsystem occultations of the Uranian $\alpha$ and $\beta$ rings exhibit quasi-periodic optical depth variations with radial wavelengths that vary with longitude. These patterns may be wakes from small moonlets orbiting exterior to these rings. Based on the observed structures in the rings, we estimate that the moonlets would need to be located $\sim100$ km exterior to the rings' semimajor axes ($106^{+22}_{-12}$ km for $\alpha$ and $77^{+8}_{-4}$ km for $\beta$) and be $2-7$ km in radius. Such moonlets could help keep the rings confined. Due to their small radii and presumed low albedo, the expected brightness of these moonlets is on the order of the noise in \textit{Voyager 2} images.
\end{abstract}


\section{Introduction}

The Uranian ring system was the second to be discovered in our solar system, thanks to multiple ground-based stellar occultation observations of Uranus on 1977 March 10 \citep{1977IAUC.3051....1E}. The nine classical rings of Uranus (named $6, 5, 4, \alpha, \beta, \eta, \gamma, \delta$, and $\epsilon$) are narrow ringlets with widths between 1 and 100 km \citep{1986Icar...67..134F}. These rings eluded discovery for so long both because they are narrow and because they are composed of extremely dark particles with geometric albedos around $0.05$ \citep{2001Icar..151...51K}. Many of the rings are also eccentric, with radial deviations from circular as large as several tens to hundreds of kilometers, and inclined by as much as $0.06^\circ$ \citep{1988Icar...73..349F}. Various theories have been proposed to explain the overall architecture of this ring system, which are best summarized in \citet{1984prin.conf...25E} and \citet{1991uran.book..327F}. The proposed solutions include a system of shepherd satellites with appropriate resonances at each ring edge, as well as embedded satellites within the rings \citep{1979AJ.....84.1225D}.

While the ground-based occultation data obtained since 1977 have provided the rings' orbital elements and widths, they do not have sufficient resolution to reveal the rings' fine-scale interior structure. The only data regarding this structure come from the three occultation experiments performed during \textit{Voyager 2's} flyby of Uranus in January of 1986. Details regarding the \textit{Voyager 2} occultations are found in \citet{1987AJ.....94..178H} (Ultraviolet spectrometer---UVS),  \citet{1989Icar...78..131G} (Radio Science Subsystem---RSS) and \citet{1990Icar...83..102C} (Photopolarimeter Subsystem---PPS). These high-resolution observations enabled the structure of these narrow rings to be examined in more detail. For example, \citet{1987AJ.....93..724P} used the orbital elements determined by \citet{1987AJ.....93.1268O} for the small moons Cordelia and Ophelia and the orbital elements of the $\epsilon$ ring \citep{1986Icar...67..134F} to show that the moons' eccentric resonances located on the inner and outer edges of the $\epsilon$ ring could be keeping the ring radially confined. They also showed that other resonances with these moons could confine the outer edges of the $\delta$ and $\gamma$ rings. Forty years after the rings' discovery, these two shepherd satellites remain the best evidence for moonlets confining narrow rings. However, the search for additional moons to shepherd the remaining ring edges has not been successful \citep{1990Natur.348..499M}.

Here we will use the high-resolution \textit{Voyager} data to analyze the interior structure of the $\alpha$ and $\beta$ rings, both of which exhibit quasi-periodic optical depth variations. These structures are unusual in that their radial wavelengths vary with longitude, even after accounting for the observable changes in the ringlet's width. Such longitudinally variable wavelengths are a characteristic of wakes generated by nearby moons, and so we explore that possibility in depth in this paper. \citet{1986Icar...66..297S} developed a model of such moonlet wakes to determine the location of the small moonlet Pan in the Encke Gap of Saturn's A ring using occultation data. We use their model to determine the possible location of a tiny moonlet just exterior to the $\beta$ ring and of another moonlet possibly perturbing the $\alpha$ ring. These moons have locations and masses that are consistent with existing limits and dynamical constraints, and they could help confine these rings.

In Section 2 we describe the \textit{Voyager 2} RSS occultations used for this analysis. In Section 3 we summarize the theoretical background of the narrow ring problem and moonlet wakes. Section 4 shows the application and results of our ring occultation scan analysis and the resulting estimates of the $\alpha$ and $\beta$ moonlet orbits and masses. In Section 5 we show that these moonlets could plausibly have avoided detection in the \textit{Voyager 2} images. Lastly, we present our discussion and conclusions in Section 6. Appendix \ref{appendixA} describes our method of calculating the RSS $\alpha$ ingress wavelength, while Appendix \ref{appendixB} discuses an analysis of the PPS occultations.

\section{Occultation Data}

\begin{figure*}
\includegraphics[width=0.95\linewidth]{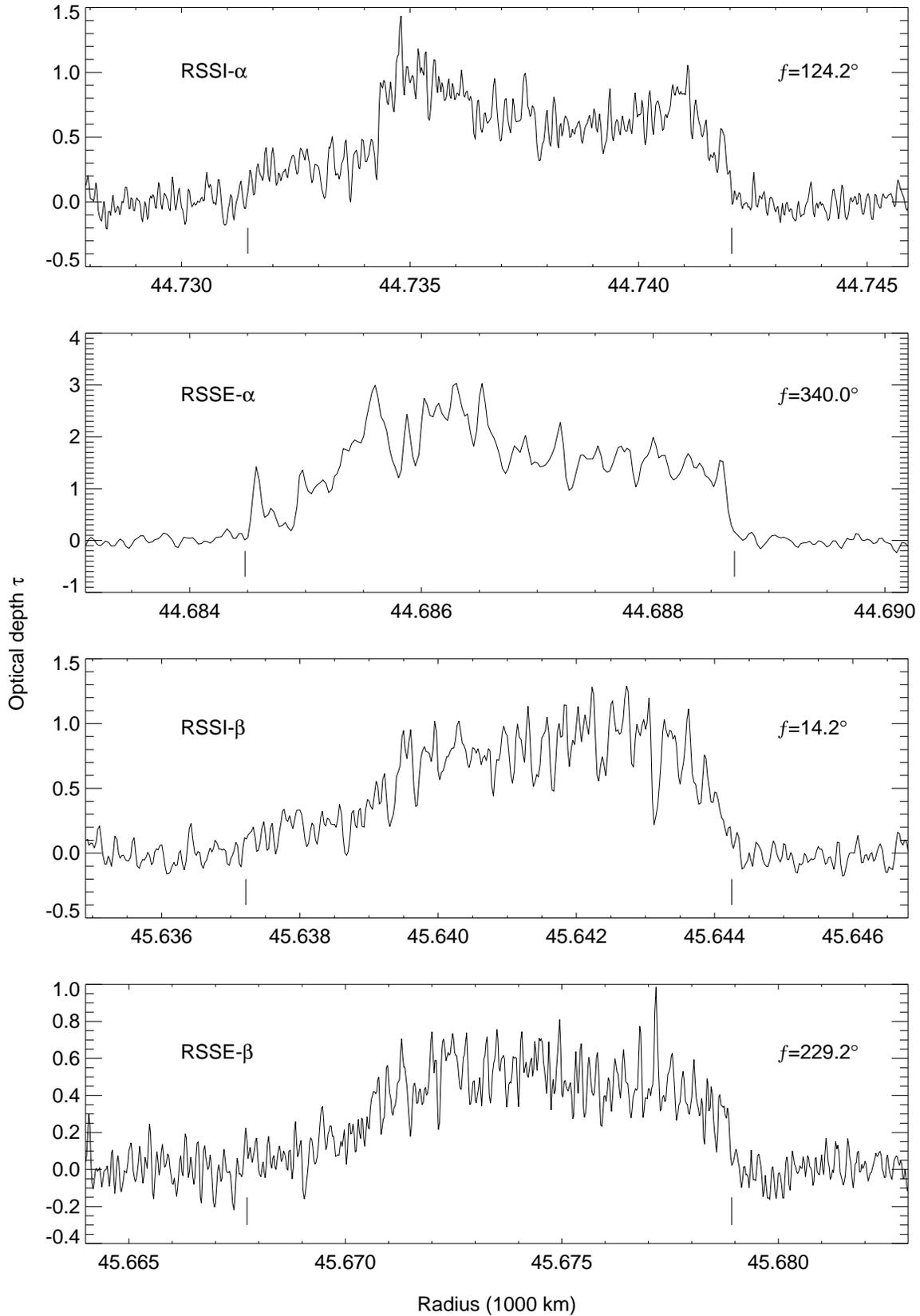}
\centering
\caption{RSS occultation scans of the $\alpha$ and $\beta$ rings. Note the varying scales of both optical depth and radius for each plot. The plots are stretched with these varying scales to a common width to allow comparison of features in their radial structure. This stretching is most apparent in the RSSE-$\alpha$ scan, whose actual width is only 4.22 km, compared to the RSSE-$\beta$ scan, whose actual width is 11.19 km. Ring edge radii from Table \ref{occ. geom.} are marked with single dashes below the data. The true anomaly of each scan is provided in the upper right. The two $\beta$ scans have different wavelengths of quasi-periodic optical depth variations near their outer edges. The $\alpha$ egress scan also shows a periodic structure near its outer edge, and some narrow evenly spaced dips are in the ingress scan. \label{ring occ.'s}}
\end{figure*}

\floattable
\begin{deluxetable}{cccccccc}
\tablecaption{Geometry of Radio Science Subsystem (RSS) occultations\label{occ. geom.}}
\tablewidth{0pt}
\tablehead{
\colhead{Ring} & 
\colhead{Occ.} & 
\colhead{True Anom.} & 
\colhead{Mid-time} & 
\colhead{Mid-rad.} & 
\colhead{Mid-long.} & 
\colhead{Inner} & 
\colhead{Outer} \\
\colhead{} & 
\colhead{} & 
\colhead{(deg)} & 
\colhead{(hr:min:s)} & 
\colhead{(km)} & 
\colhead{(deg)} & 
\colhead{Edge (km)} & 
\colhead{Edge (km)}
}
\startdata
$\alpha$ & RSSI & 124.4 & 19:55:58.385 & 44,736.75 & 342.1 &
44,731.45 & 44,742.04\\
& RSSE & 340.2 & 22:39:26.294 & 44,686.59 & 198.1 & 44,684.48 & 44,688.70\\
$\beta$ & RSSI & 13.9 & 19:54:07.459 & 45,640.74 & 342.5 & 45,637.22 & 45,644.25\\
& RSSE & 228.9 & 22:41:26.104 & 45,673.33 & 197.7 & 45,667.73 & 45,678.92\\
\enddata
\tablecomments{The appended labels of I and E stand for ingress and egress. The true anomaly of each ring at the time of their respective ring intercept mid-times is calculated from the inertial longitudes provided and updated precession rates provided by R. G. French and summarized in Table \ref{orb elements}. Mid-times listed are the times of ring intercept measured in seconds after UTC 1986 January 24 00:00:00, when the RSS microwaves intercepted the mid-ring radius. The corresponding mid-radii are explicitly calculated as halfway between the two edge radii and may differ slightly from those of other sources whose mid-radii refer to a location weighted on the equivalent depth of the occultations scans. Mid-longitudes similarly correspond to mid-ring radii and mid-times.} 
\end{deluxetable}

The data for this investigation consist of occultations obtained by the RSS instrument on board the \textit{Voyager 2} spacecraft, available on NASA's Planetary Data System Ring-Moon Systems Node website.\footnote{http://pds-rings.seti.org} RSS generated a complete ingress and egress occultation of the rings (separated longitudinally by $\sim145^{\circ}$) by transmitting microwave radiation of wavelengths 3.6 cm (X band) and 13 cm (S band) through the rings to ground stations on Earth \citep{1986Sci...233...79T,1989Icar...78..131G}. These radio wavelengths are not to be confused with the ring density wavelength of the rings' wake structure to follow. Note that the PPS stellar occultation of $\beta$ Persei (Algol) is of higher spatial resolution but lower signal-to-noise ratio than the RSS occultations \citep{1990Icar...83..102C,1995AJ....109.2262G}, and the UVS performed the same occultations as the PPS, but at a lower resolution. 
We find that these stellar occultations do not have sufficient signal-to-noise to provide further evidence for or against the idea that these rings may contain moonlet wakes. Our analysis of the PPS occultations is included in Appendix \ref{appendixB}. 

Table \ref{occ. geom.} provides a summary of the RSS occultation data set used here, giving mid-times, inertial longitudes and radii, and true anomalies for the $\alpha$ and $\beta$ rings at the time of the ingress and egress occultations, as well as our estimated positions of the ring edges, consistent with \citet{1989Icar...78..131G}. Figure \ref{ring occ.'s} shows both the ingress and egress RSS occultations of the $\alpha$ and $\beta$ rings \citep[for occultation scans of all rings with all instruments see][]{1991uran.book..327F}. We show the ring profiles scaled so that the rings appear to have a common width. This provides a better view of the internal structure of the rings and enables direct comparisons of the ring scans with different true anomalies and widths. The outer region of each ring contains a series of dips and peaks, most obviously seen in the $\alpha$ and $\beta$ egress scans. In both ringlets, these periodic structures do not have the same wavelength in the ingress and egress scans. Such longitudinally variable wavelengths are atypical of many ring features, like density waves, but are characteristic of moonlet wakes, and so we hypothesize that these structures are caused by nearby perturbing moonlets.

\section{Theoretical Background}

\begin{figure*}
\includegraphics{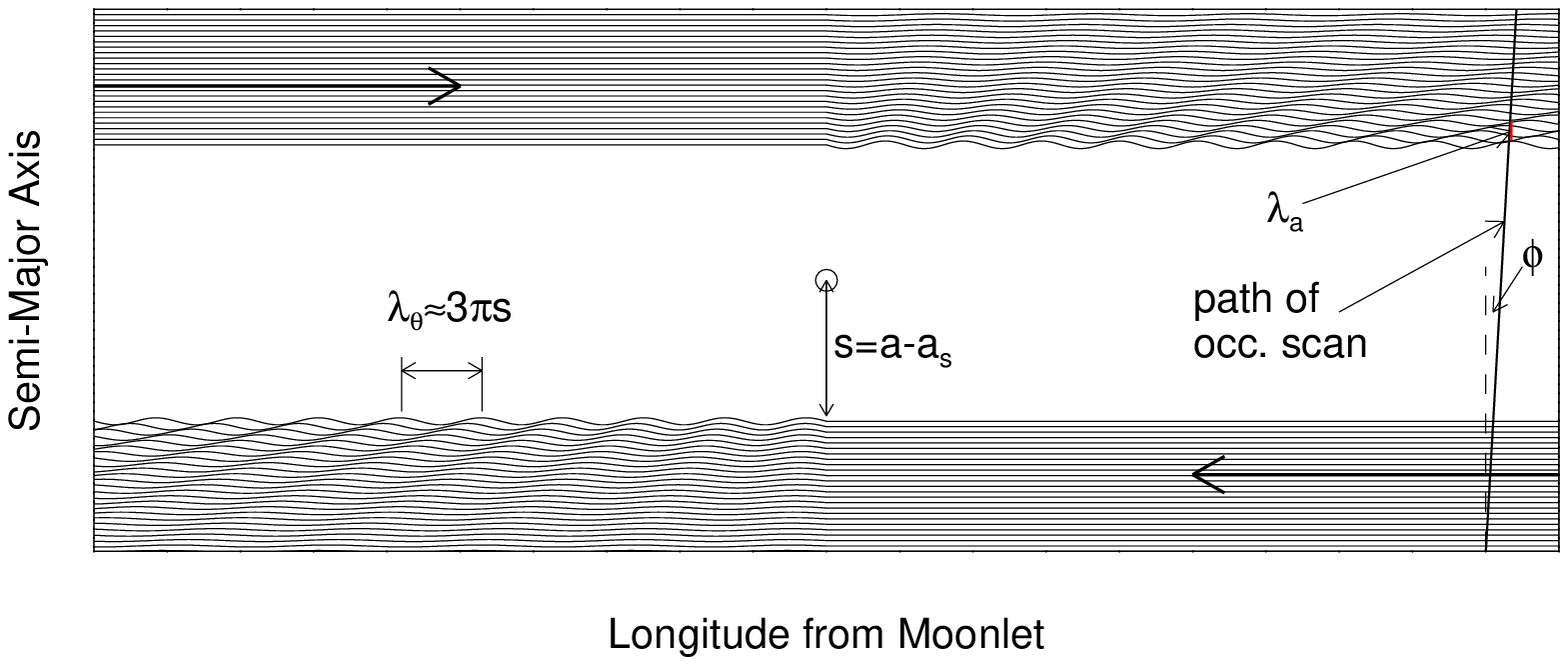}
\centering
\caption{Schematic of the ring--moon wake interaction, in a reference frame centered on the moon. Ring material closer to the planet in the lower half of the figure is moving faster than the moon and passes it to the left, and the ring material in the upper half is moving slower and is passed by the moon. The diagonal line on the right side is meant to simulate the path of an occultation scan through the rings. \label{wake schematic}}
\end{figure*}

\floattable
\begin{deluxetable}{ccccccc}
\tablecaption{Ring orbital elements \label{orb elements}}
\tablewidth{0pt}
\tablehead{
\colhead{Ring} & 
\colhead{$a$ (km)\tablenotemark{a}} & 
\colhead{$ae$ (km)\tablenotemark{a}} & 
\colhead{$\varpi_0$ $(deg)$\tablenotemark{a}} &
\colhead{$\dot{\varpi}$ $(deg/$day$)$\tablenotemark{a}} &
\colhead{$q_e$\tablenotemark{b}} &
\colhead{$q_{\varpi}$\tablenotemark{b}}
} 
\startdata
$\alpha$ & $44,718.96\pm0.13$ & $34.01\pm0.10$ & $332.75\pm0.35$ & $2.18542\pm0.00009$ & 0.57 & $+0.07$ \\
$\beta$ & $45,661.39\pm0.11$ & $20.15\pm0.09$ & $223.30\pm0.55$ & $2.03115\pm0.00012$ & 0.27 & $+0.003$ \\
\enddata
\tablenotetext{a}{Provided by R. G. French. Listed longitude of pericenter, $\varpi_0$, corresponds to the epoch: UTC 1977 March 10, 20:00:00.}
\tablenotetext{b}{$q_e=a\delta e/\delta a + e$ and $q_{\varpi}=ae\delta \varpi/\delta a$ are from Table VII of \citet{1991uran.book..327F} and are determined only from the \textit{Voyager 2} RSS data.}
\end{deluxetable}

Over time, the continuous dissipation of energy through inelastic particle collisions in a dense ring will cause it to spread out radially \citep{1982ARA&amp;A..20..249G,1984prin.conf..447S}. In standard models, an unperturbed, narrow ringlet should spread on timescales of only a small fraction of the age of the solar system, $\sim2500$ yr for rings comparable to the $\alpha$ and $\beta$ rings \citep[p. 497]{1999ssd..book.....M}. One possible mechanism for confining ring edges is through the gravitational perturbations of a nearby satellite \citep{1982ARA&amp;A..20..249G,1984prin.conf..713B}. Such moons should produce observable structures in a ring in the form of wavy edges and moonlet wakes. Indeed, \citet{1991Natur.351..709S} was able to find Saturn's small moon Pan in \textit{Voyager 2} images after determining its orbital elements based on optical depth variations observed in occultation scans of the surrounding A ring \citep{1986Icar...66..297S}.

To understand this mechanism, consider a ring particle on a circular orbit at semimajor axis, $a$, slowly passing by a moonlet at semimajor axis $a_s > a$. While passing by the moonlet, the ring particle gains a net gravitational acceleration in the radial direction toward the moonlet and thus a small component of velocity in the radial direction, $v_r$. This radial velocity puts the ring particle on a slightly eccentric orbit with an apoapsis located one-quarter of its orbit downstream from the moonlet--ring interaction. As additional ring particles from the original circular ringlet pass by the moonlet, they undergo the same interaction. The gradually shifting apoapsis location induced in the ring particle orbits forms a wavy edge to the ring with azimuthal wavelength, $\lambda_{\theta}(a)\approx3\pi\left|s\right|$ \citep{1985ApJ...292..276C}, where $s=a_s-a$. The wavy edge appears ahead of the moonlet if the ring is interior to the moon and trails the moonlet if the ring is exterior (see Figure \ref{wake schematic}). A moonlet a few kilometers wide would only produce variations in the edge position of around 20 m  \citep[calculated from][Equation (3)]{1985ApJ...292..276C}, which would be undetectable in the \textit{Voyager 2} data. However, the azimuthal wavelength's dependence on $s$ causes it to vary significantly over the width of a ring, with ring particle streamlines farther from the moonlet exhibiting a longer wavelength. Over a number of periods these adjacent streamlines start to go out of phase. This crowding of the streamlines results in a pattern with alternating areas of higher surface density and areas of lower surface density. We observe this as quasi-periodic optical depth variations in the occultation scans of the ring downstream from the moonlet, known as a moonlet wake.

\citet{1986Icar...66..297S} developed a model precisely for the purpose of finding the location of a moonlet given measurements of these wake wavelengths at multiple longitudes. They found the wavelength in a linear scan to be
\begin{equation} \label{eq:1}
\lambda_a \approx 3 \pi \frac{s^2}{a_s \left| \theta \right|} \left[1-\left|\frac{s}{a_s \theta} \right| \tan(\phi)\right].
\end{equation}
Here $\theta$ is the angle of azimuthal separation between the moonlet and the longitude of the occultation scan, and $\phi$ is the angle the occultation scan makes with the radial direction. This wavelength increases as $s^2$ and decreases inversely with azimuthal separation from the moonlet. We should note here that the wavelength, $\lambda_a$, is specifically referring to the wavelength of optical depth variations in a radial occultation scan of a circular ring. Because the Uranian rings are actually eccentric and their widths vary systematically with their average radius, the wavelength we measure in our occultations is not representative of the actual wavelength the moonlet would produce using this model. We therefore convert all of our occultation scans to `semimajor axis space' using the formula
\begin{equation} \label{eq:2}
da=\frac{dr}{(1-q_e\cos(f)-q_{\varpi}\sin(f))},
\end{equation}
adapted from \citet{1991uran.book..327F}, where $q_e$ and $q_{\varpi}$ are the eccentricity and pericenter gradients of the ringlet (see Table \ref{orb elements}) and $f$ is the true anomaly (see Table \ref{occ. geom.}). This allows us to translate the observed optical depth profiles into semimajor axis space and thus compute the appropriate wavelength of the wake $\lambda_a$. If $\phi$ is small, as it is for the RSS occultations  \citep[see Figure 2 of][]{1991uran.book..327F}, we can neglect the second term on the right-hand side of Equation \ref{eq:1} and solve for $a_s$ as a function of $\theta$:
\begin{equation} \label{eq:3}
a_s \approx a + \frac{\lambda_a \left| \theta \right|}{6 \pi} \pm \sqrt{ \frac{\lambda_a^2 \theta^2}{36 \pi^2} + \frac{a \lambda_a \left| \theta \right|}{3 \pi}}.
\end{equation}
We can therefore plot a curve $a_s(\theta)$ giving the possible locations of a moon that could be responsible for producing an observed value of $\lambda_a$. These positions can further be expressed in terms of an absolute longitude of the satellite in an inertial reference frame ($\theta_s$), by subtracting the appropriate longitude of the occultation data. We compare the location curves of multiple scans (after shifting their longitudes to a common epoch) and look for a common location as cause of the optical depth variations. Based on the location and amplitude of the optical depth variations, we can determine the mass of the putative moonlet (see Section 4 below).  

\section{Occultation analysis}

To determine the wavelength of the optical depth variations in each occultation scan, we use a combination of wavelet and localized Fourier transformations. Wavelet transforms provide maps of the strength of periodic structures as a function of radius and wavelength and have proven useful for studying a variety of structures in dense rings \citep[e.g.][]{2007Icar..189...14T}. The extent of the optical depth variations in the $\alpha$ and $\beta$ rings is only a few kilometers at most, so we do not expect the wavelength to vary with radius by more than $\sim$1-5\%. However, the spatial aspect provided by a wavelet transform illustrates where the periodic signal is detectable and demonstrates that it has different wavelengths in the two scans. Once the appropriate regions are identified through wavelet analysis, more precise wavelengths are determined via a Fourier transformation of the relevant ring regions. 

\begin{figure*}
\includegraphics[scale=.6,angle=90]{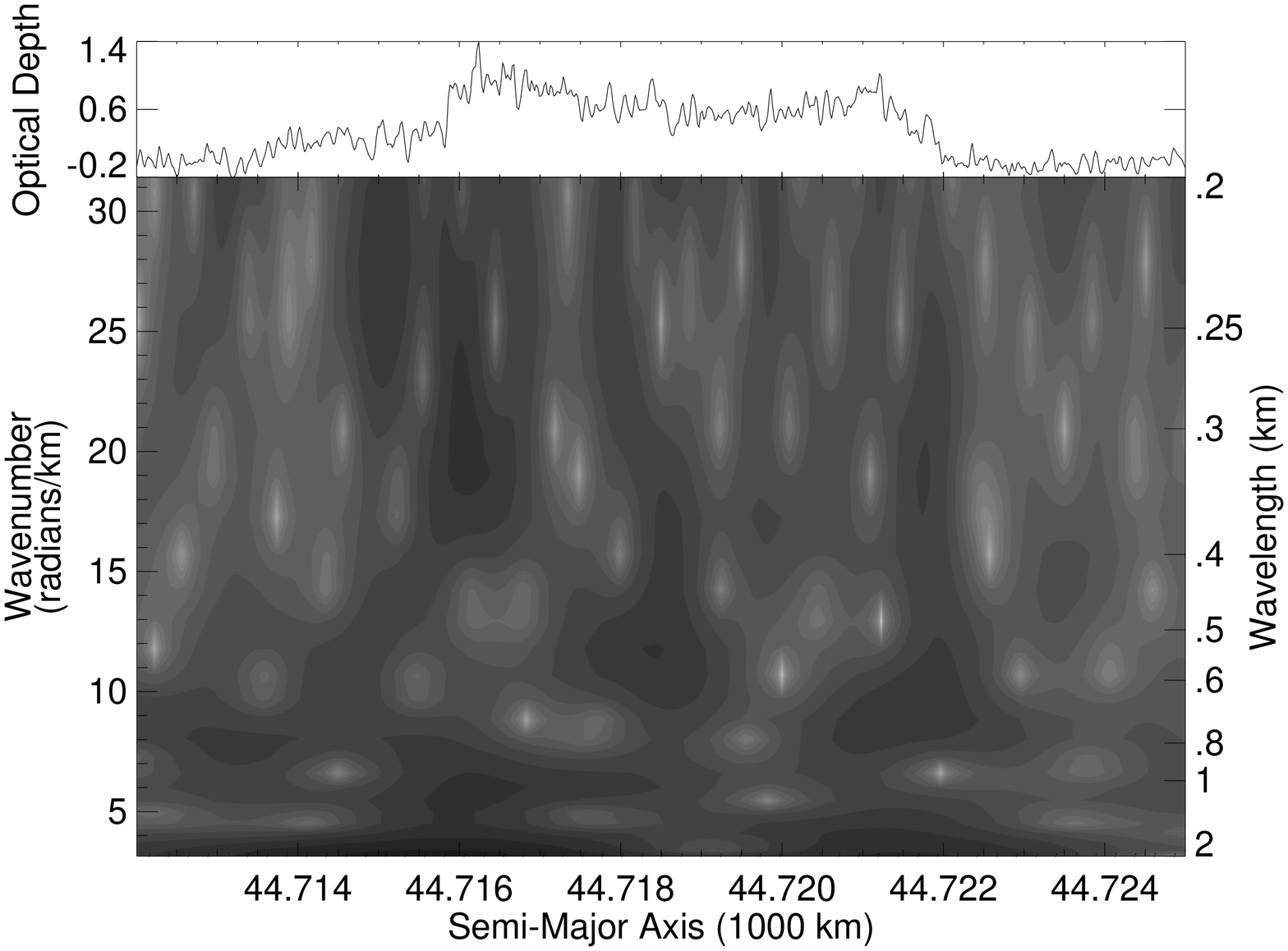}
\includegraphics[scale=.6,angle=90]{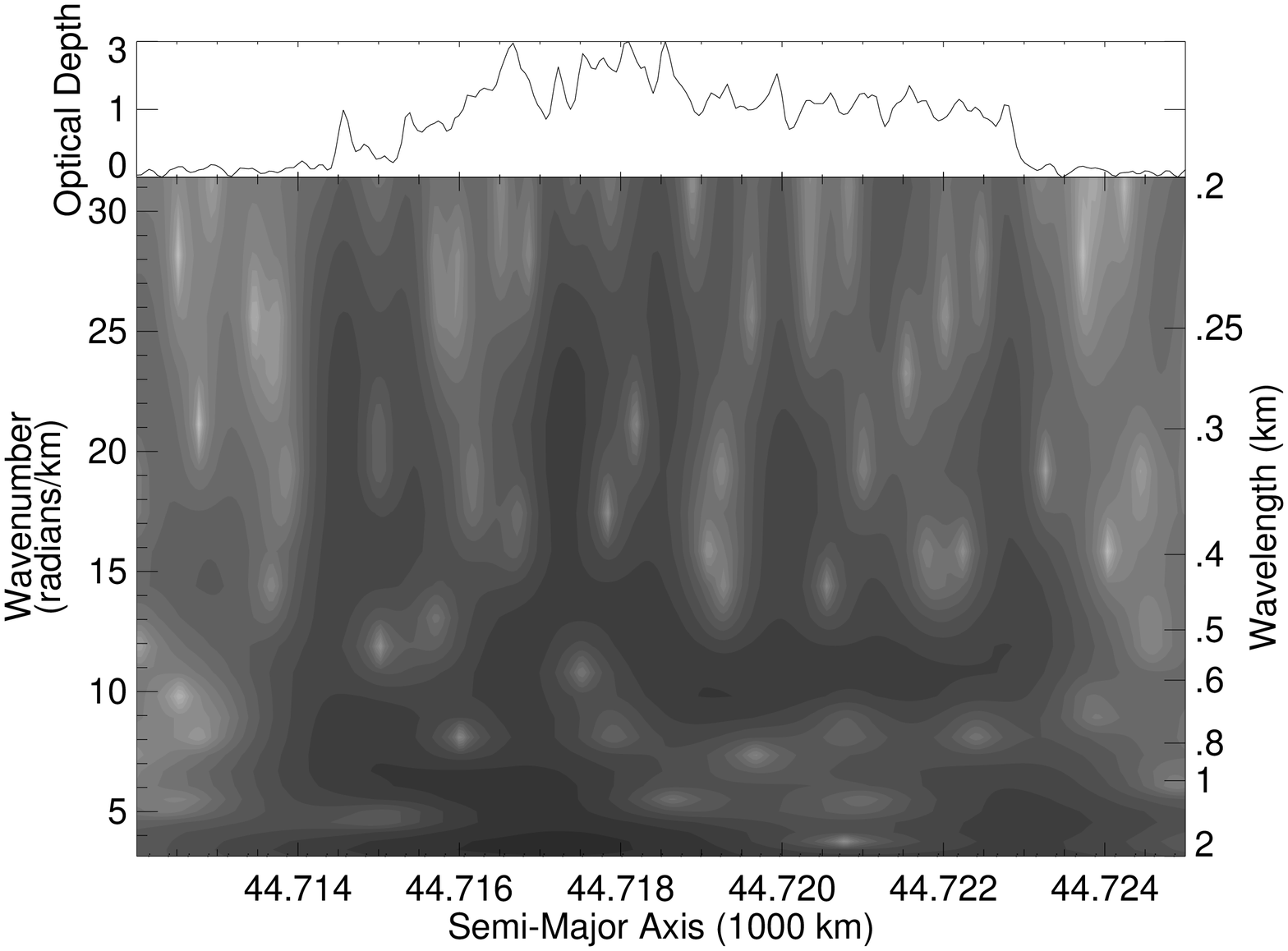}
\centering
\caption{We use wavelet transforms to determine the wavelengths of the quasi-periodic optical depth variations of the $\alpha$ ingress (top) and egress (bottom) occultation scans. The strength of the periodic signal for a given radius and wavenumber are shown in the contour map, where darker corresponds to a stronger signal. The egress wavelet shows a strong sinusoidal periodic structure from 44719-44723 km with a wavelength of 0.59 km. The ingress scan is composed of periodic sharp dips and peaks that do not produce clear signals at one wavelength in these transforms. \label{alpha wavelet}}
\end{figure*}

\begin{figure*}
\includegraphics[scale=.6,angle=90]{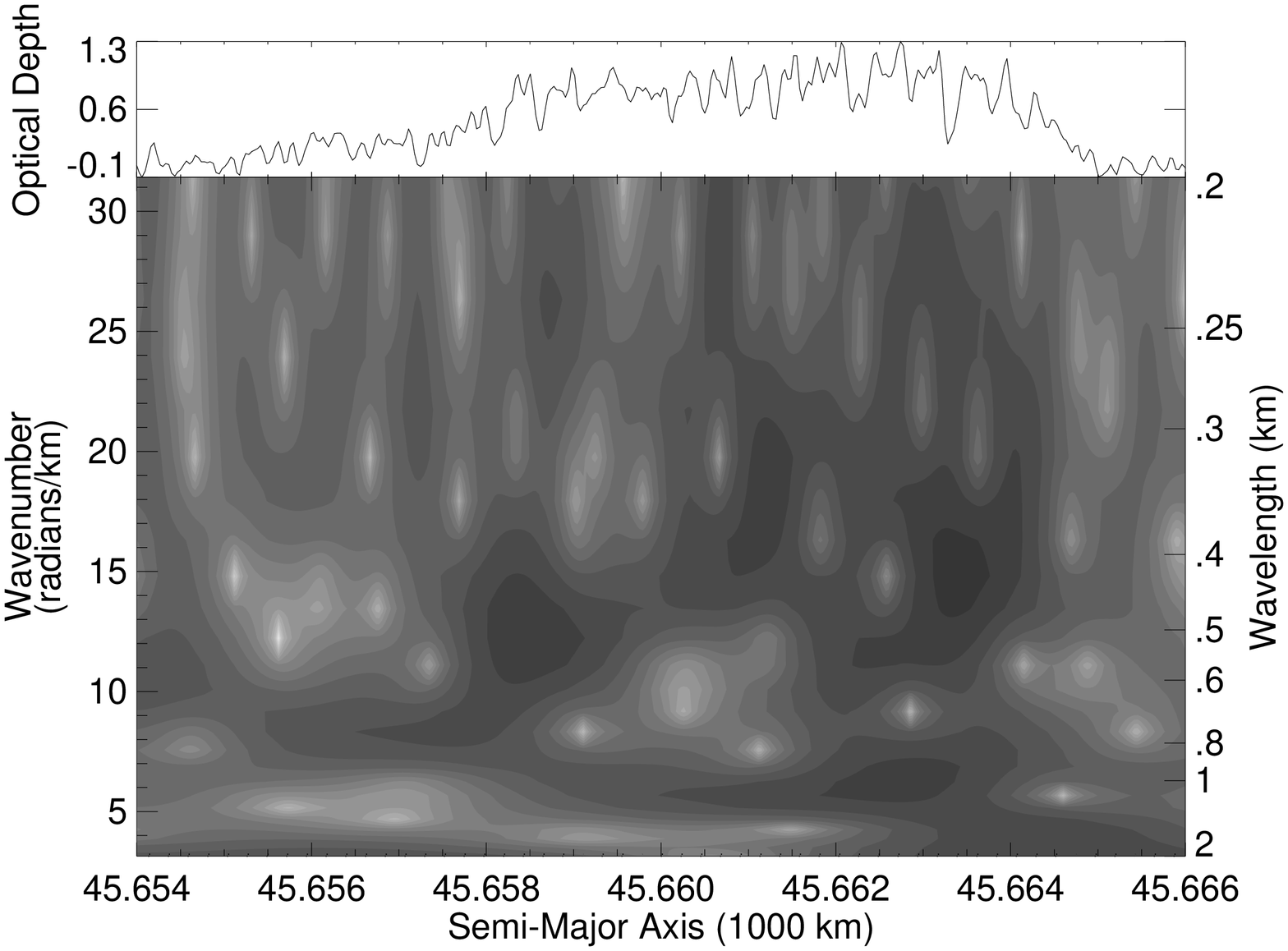}
\includegraphics[scale=.6,angle=90]{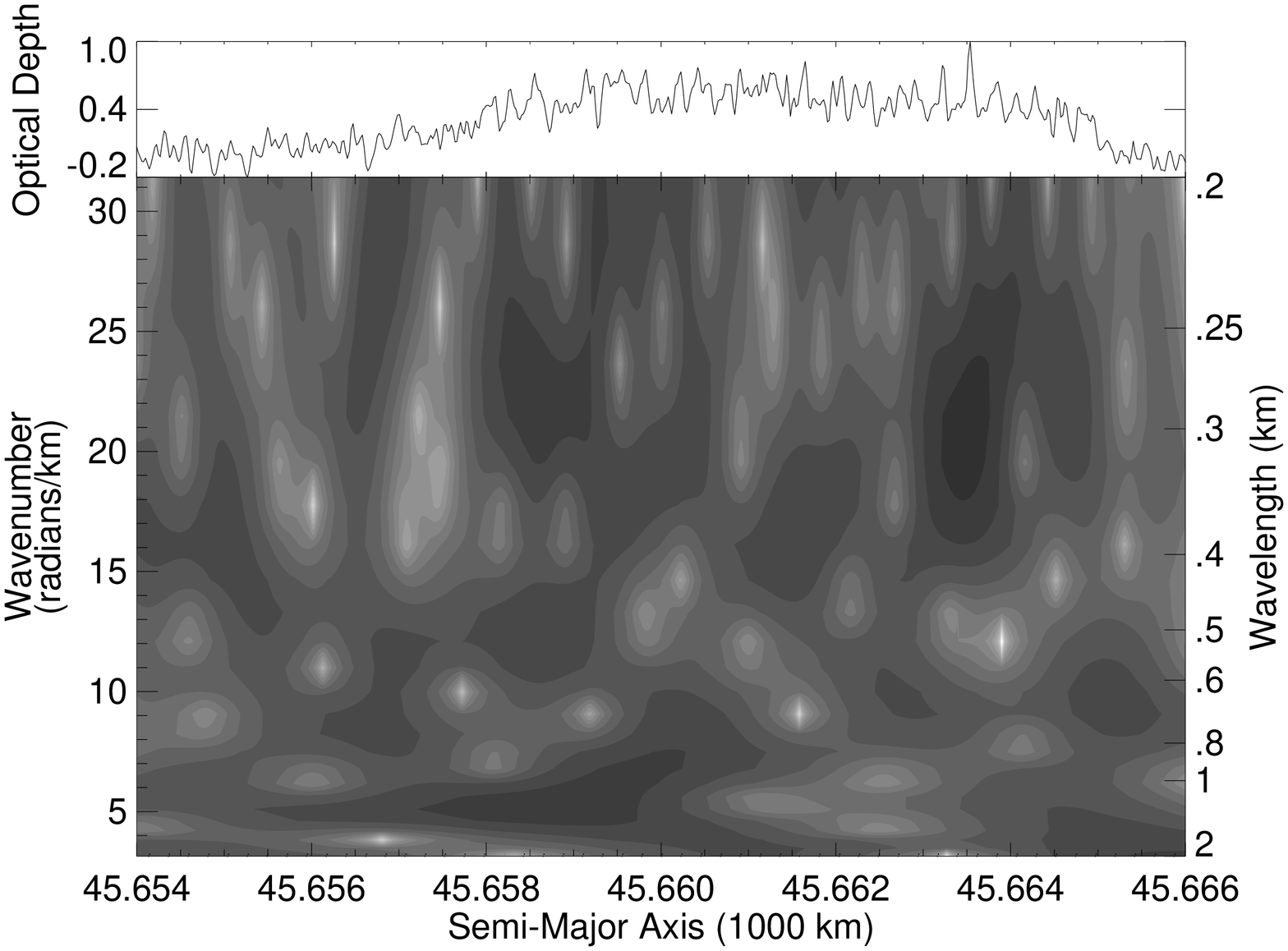}
\centering
\caption{The $\beta$ ring wavelet transforms detect different wavelengths of the optical depth variations seen near the outer edge of the ring in each scan. From 45663-45664 km we find a wavelength of 0.45 km for the ingress scan and 0.31 km for the egress scan. Fourier transforms of these radial regions are shown in Figure \ref{ft plot}. \label{beta wavelet}}
\end{figure*}

\begin{figure}
\includegraphics[width=.8\linewidth]{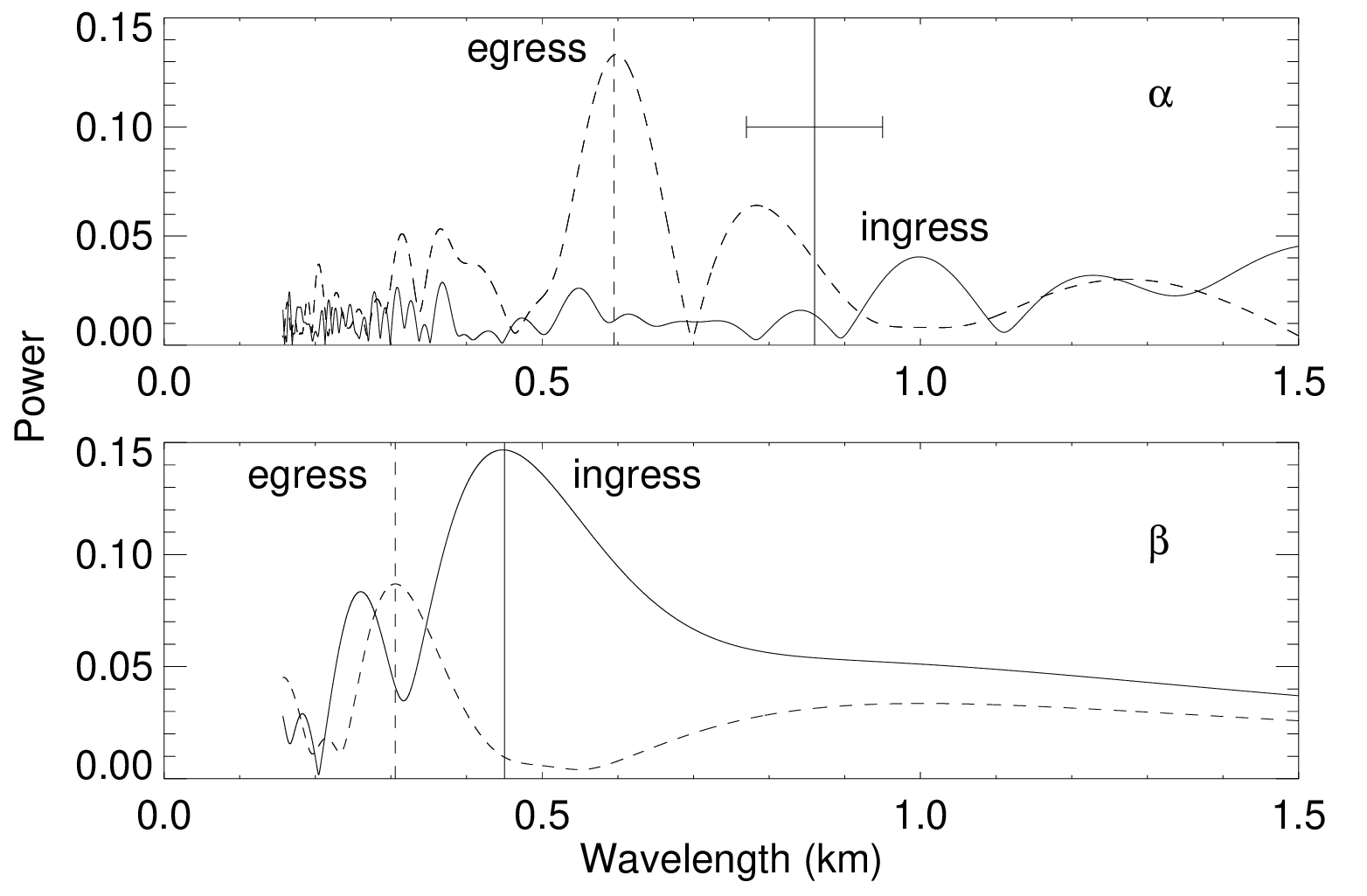}
\centering
\caption{Localized Fourier transforms of regions identified in the wavelet transforms from Figure \ref{alpha wavelet} and \ref{beta wavelet} for the $\alpha$ (top) and $\beta$ (bottom) rings. The top plot contains the transforms of the $\alpha$ ring scans from $a=44,719$ to $44,723$ km. The $\alpha$ ingress scan wavelength, shown with its larger error bar (uncertainty of other scan wavelengths is small), was not determined from this Fourier transform; see Appendix \ref{appendixA}. We did not consider the bump at $\sim1$ km in the $\alpha$ ingress Fourier spectra to be significant. If it did happen to be the real wavelength, it also produces a solution consistent with a small moonlet of about 2 km in radius. The $\beta$ ring plot contains the transform of both scans from $a=45,663$ to $45,664$ km. We plot transforms of the ingress scans with solid lines and the egress scans with dashed lines. Exact wavelength values and uncertainties are summarized in Table \ref{moon locations}. We believe that the smaller bump at $\sim0.25$ km in the $\beta$ ingress transform is a harmonic of the actual wavelength. \label{ft plot}}
\end{figure}
 
We compute the wavelet transform for each profile with the standard wavelet routine in the IDL language \citep{1998BAMS...79...61T} using a Morlet mother wavelet with $\omega_0=6$. Figures \ref{alpha wavelet} and \ref{beta wavelet} show the resulting wavelet transforms for each RSS $\alpha$ and $\beta$ profile as a function of semimajor axis and wavenumber. Here the occultations' radii have been translated to effective semimajor axes so that we can obtain the appropriate $\lambda_a$ wavelengths. The darker regions show the locations where the periodic signal is strongest. The most obvious periodic pattern is seen in the outer half of the $\alpha$ ring's egress scan between semimajor axes of 44,719 to 44,723 km, where the wavelength is $0.59$ km. The $\beta$ ring wavelets show similar periodic optical depth variations near the ring's outer edge (semimajor axes between 45,663-45,664 km) with different wavelengths in each scan, 0.45 km for ingress and 0.31 km for egress. In order to obtain precise estimates of the patterns' wavelengths, we compute over-resolved Fourier transforms of the above regions, which are shown in Figure \ref{ft plot}. Note that we actually determined the wavelengths from gaussian peak fits of the Fourier power versus wavenumber and then converted to wavelengths. This was done because the peaks in the power spectrum are more symmetric in wavenumber space. The wavelengths derived from these methods are given in Table \ref{moon locations}, along with their uncertainties $\sigma=\sqrt{2}\delta r/N$ \citep{2007Icar..188...89H}, where $\delta r$ is the occultation scan's radial resolution and $N$ is the number of wavelengths that extend across the radial region where the wavelength is measured.

Compared with the other occultations, the wavelet transform of the $\alpha$ ring ingress profile shows much more disorganized signals. This is most likely because the periodic structure in this profile seems to consist of periodic narrow dips, rather than a sinusoidal wave. The variations in the morphology of the $\alpha$ ring patterns are similar to those seen in a density wave located within the Maxwell ringlet in Saturn's rings. Indeed, \citet{2016Icar..279...62F} found that the detailed morphology of the optical depth variations associated with this wave varied systematically with the ringlet's true anomaly. While at many true anomalies the optical depth variations were sinusoidal, when the true anomaly was close to $90^\circ$ (i.e. similar to the $\alpha$ ingress scan), the optical depth variations in the wave become very narrow dips and peaks, similar to those seen in the $\alpha$ ingress profile. The lack of a sinusoidal periodic structure made Fourier-transform-based estimates of the pattern wavelength problematic, and so we found it more effective to estimate the wavelength of the $\alpha$ ingress scan by visual inspection and determination of the separation of the individual dips in optical depth (see Appendix \ref{appendixA}). 

\begin{figure*}
\includegraphics[width=.85\linewidth]{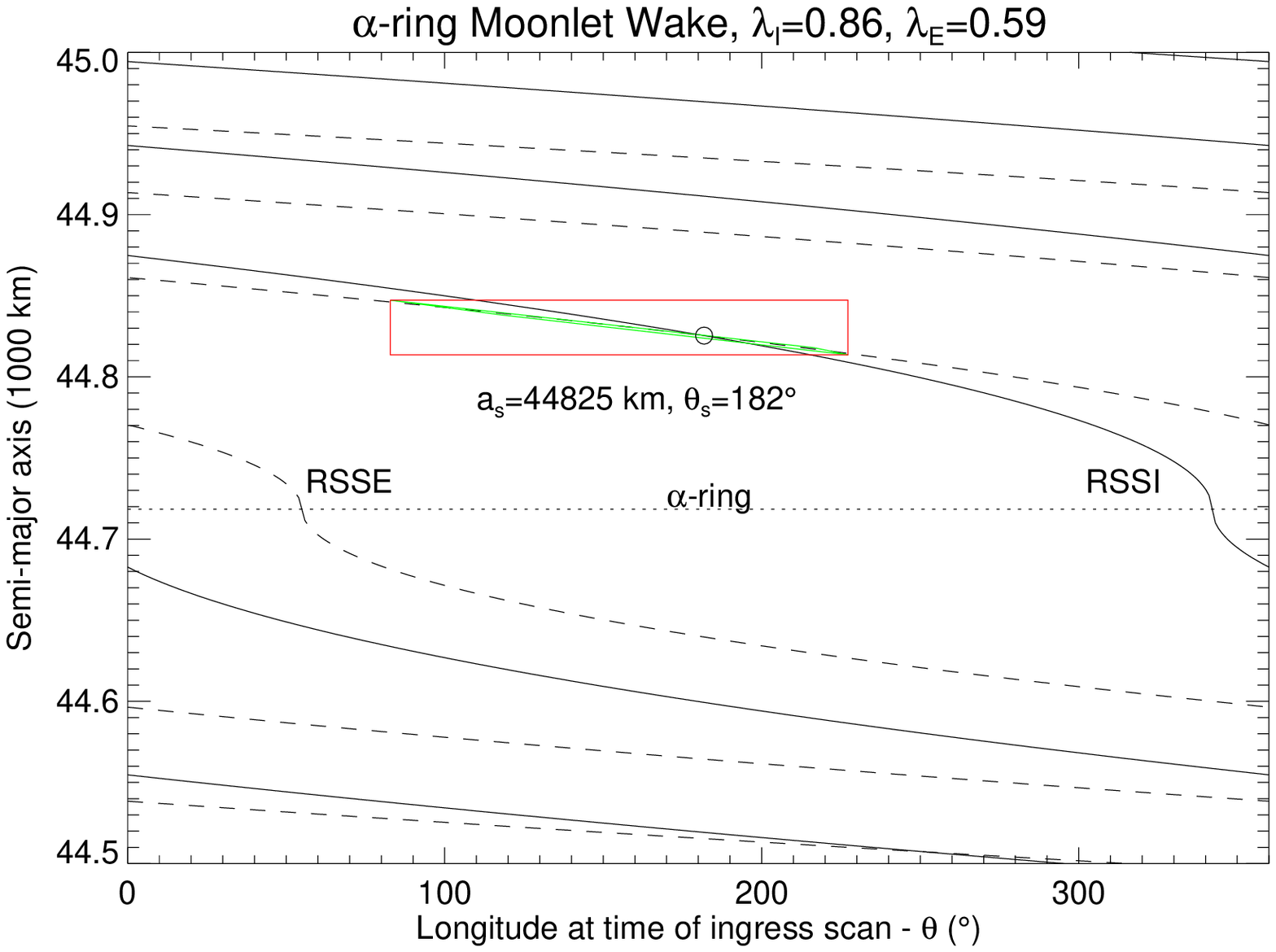}
\includegraphics[width=.85\linewidth]{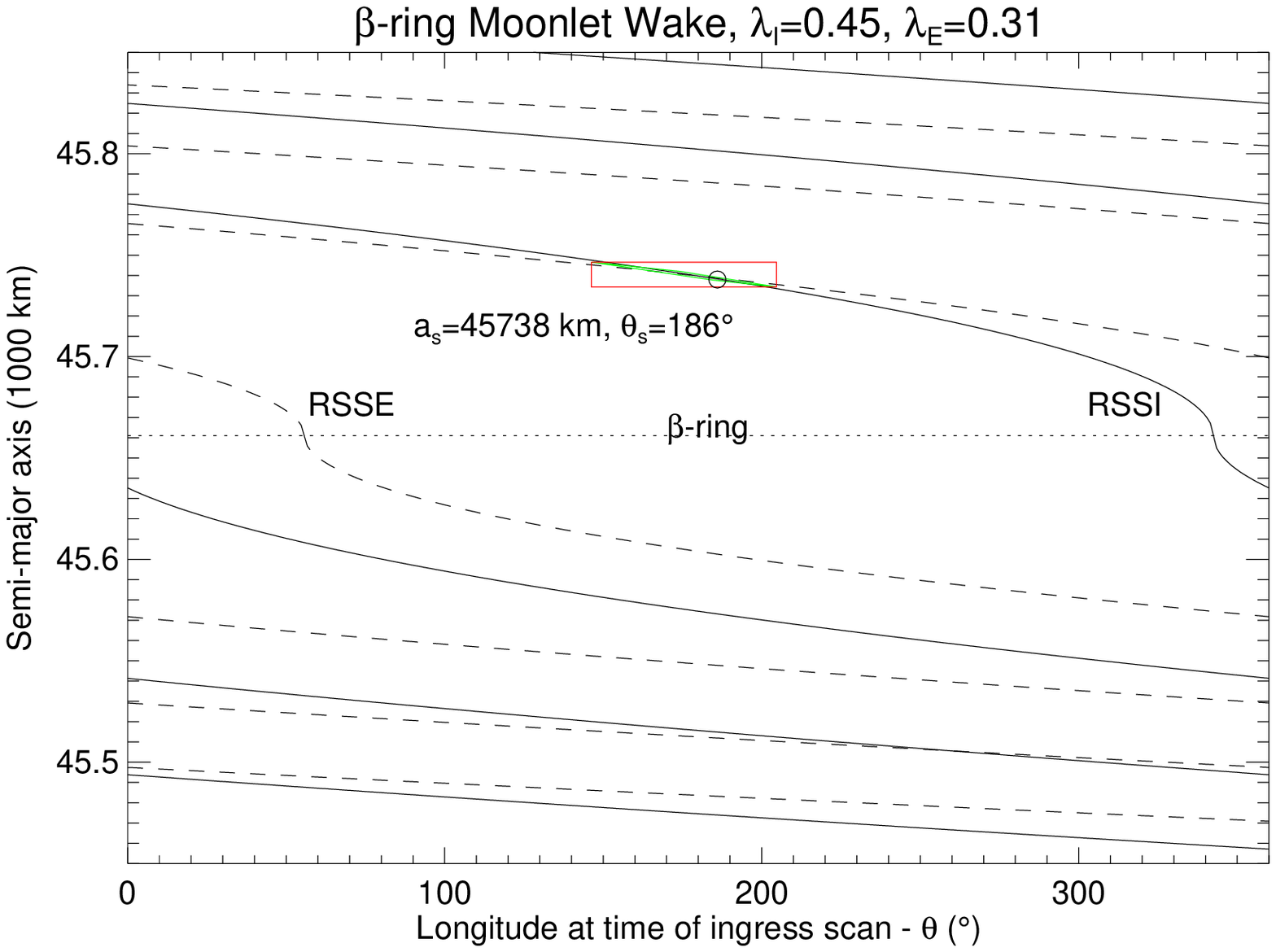}
\centering
\caption{Location of the $\alpha$ (top) and $\beta$ (bottom) moonlet determined by plotting the curves of Equation \ref{eq:3} shifted into the time frame of the RSS ingress occultation with the ingress and egress wavelengths of Table \ref{moon locations}. A moonlet located where the ingress (solid) and egress (dashed) curves cross, labeled with a circle, would be able to produce the optical depth variations seen in both scans. The red box outlines the maximum extent of the uncertainties in $a$ and $\theta$, although this exaggerates the actual range in uncertainties for the location, shown as the very narrow green parallelogram-shaped area resulting from calculations of all combinations of wavelength uncertainties. If we consider the bump in the Fourier spectrum of the $\alpha$ ingress scan at $\sim1$ km, the location where the curves cross, as a result of this larger wavelength, moves to a smaller semimajor axis, resulting in a moonlet slightly closer to the ring with a smaller radius ($\sim2$ km).  \label{wake plot}}
\end{figure*}

We insert the derived wavelengths, along with the rings' most precise semimajor axes contained in Table \ref{orb elements}, into Equation \ref{eq:3} to generate the curves of allowed locations of moons shown in Figure \ref{wake plot}. In each panel, both curves are shown in a reference frame computed at the time of the ingress scan. The location where the ingress (solid lines) and egress (dashed lines) curves cross gives the semimajor axis $a_s$ and the inertial longitude $\theta_s$ a moonlet would need to have in order to cause the optical depth variations seen in both scans. In Table \ref{moon locations}, we show the input wavelengths, $\lambda_a$, for the ingress and egress scans and moonlet locations (circles in Figure \ref{wake plot}) consistent with both scans for the $\alpha$ and $\beta$ rings. The uncertainties in the wavelengths propagate through the calculations of the moonlet locations, whose listed uncertainties are then taken from the largest deviations in each case. We find for both rings that a moonlet located about 100 km exterior to each ring could cause the optical depth variations seen in their occultation scans. This puts the moonlets orbiting outside the maximum radial extent of the rings due to their eccentricities (see Table \ref{orb elements}), which are physically sensible locations. 

\floattable
\begin{deluxetable}{ccccccccc}
\centering
\tablecaption{Moonlet Locations \label{moon locations}} 
\tablehead{
\colhead{Ring} & \colhead{$\lambda_{I}$ (km)} & \colhead{$\lambda_{E}$ (km)} & \colhead{$a_s$ (km)} & \colhead{$\theta_s (^{\circ})$} & \colhead{$s$ (km)}
}
\startdata
$\alpha$ & $0.86\pm0.09$\tablenotemark{a} & $0.59\pm0.01$ & $44825^{+22}_{-12}$ & $182^{+45}_{-99}$ & $106^{+22}_{-12}$ \\
$\beta$ & $0.45\pm0.01$ & $0.31\pm0.01$ & $45738^{+8}_{-4}$ & $186^{+18}_{-40}$ & $77^{+8}_{-4}$ \\
\enddata
\tablenotetext{a}{The wavelength for the $\alpha$ ingress scan was determined by visual inspection of the series of dips rather than through wavelet and Fourier analysis. Its larger error is the standard error of the mean wavelength calculated in Appendix \ref{appendixA}.}
\tablecomments{Moonlet semimajor axis, $a_s$, inertial longitude at the epoch of the ingress occultation scan (see Table \ref{occ. geom.}), $\theta_s$, and ring--moon radial separation, $s$, relative to the rings' semimajor axes in Table \ref{orb elements}, determined using the ingress and egress wavelengths, $\lambda_{I/E}$, of the quasi-periodic optical depth variations near the outer edges of the rings ($\lambda_a$).}
\end{deluxetable}

\floattable
\begin{deluxetable}{ccccccccc}
\tablecaption{Moonlet masses and radii \label{moon masses}}
\tablewidth{0pt}
\tablehead{
\colhead{Ring} & \colhead{$\tau_{peakI}$} & \colhead{$\tau_{0I}$} & \colhead{$\tau_{peakE}$} & \colhead{$\tau_{0E}$} & \colhead{$M_{sI}$ (kg)\tablenotemark{a}} & \colhead{$M_{sE}$ (kg)\tablenotemark{a}} & \colhead{$R_{sI}$ (km)\tablenotemark{b}} & \colhead{$R_{sE}$ (km)\tablenotemark{b}}
}
\startdata
$\alpha$ & 0.51\tablenotemark{c} &  0.72 & 1.88 & 1.52 & $(3^{+4}_{-2}) \times 10^{14}$ & $(1.0^{+18}_{-0.6}) \times 10^{14}$ & $4\pm1$ & $3^{+4}_{-1}$ \\
$\beta$ & 1.16 & 0.83 & 0.70 & 0.45 & $(0.5^{+0.3}_{-0.2}) \times 10^{14}$ & $(0.7^{+0.4}_{-0.2}) \times 10^{14}$ & $2.1^{+0.4}_{-0.2}$ & $2.4^{+0.6}_{-0.3}$ \\
\enddata
\tablenotetext{a}{Mass uncertainties are the extremes resulting from all combinations of input locations and their uncertainties. Note that the asymmetry of these error bars is due to the factor of $s^4$ in Equation \ref{eq:6}.}
\tablenotetext{b}{Radius uncertainties as above with uncertain mass inputs.}
\tablenotetext{c}{$\tau_{peakI}$ is less than $\tau_{0I}$ for $\alpha$ because in the case of the $\alpha$ ingress occultation we used the optical depth dips instead of peaks; thus, the $\cos(\eta \theta)$ term in Equation \ref{eq:4} goes to -1.}
\tablecomments{Moonlet masses for each ring are determined from the optical depth variations and longitudinal separations of the ingress and egress occultation scans using Equation \ref{eq:6}. Moonlet radii are calculated from the average mass of the ingress and egress scans using an estimated density of $1.3$ g/cm$^3$.}
\end{deluxetable}

We can estimate the mass of the moonlets based on the amplitude of the wakes using Equation (2) from \citet{1996Icar..124..663H},
\begin{equation} \label{eq:4}
\tau(r,\theta)=\frac{\tau_0(a)}{1+2.24j\mu\eta_0\theta\cos(\eta\theta)(a_s/s)^4},
\end{equation}
where $\tau$ is the observed ring optical depth, $\tau_0$ is the average (unperturbed) optical depth, $j$ is $+1$ for a moonlet interior to the ring and $-1$ otherwise, $\mu$ is the ratio of the satellite and planet masses, and $\eta_0$ and $\eta$ are azimuthal wave numbers defined in \citet{1986Icar...66..297S} such that
\begin{equation} \label{eq:5}
\eta=\frac{\eta_0 a_s}{s}+\eta_1+\dots=\frac{-\frac{2}{3}+\frac{5}{3} J_2 \Big( \frac{r_p}{a_s} \Big)^2+\dots}{s/a_s}+\frac{1}{6}+\dots
\end{equation}
The mass of the satellite can therefore be calculated as
\begin{equation} \label{eq:6}
M_s=\frac{M_p(\tau_0/\tau_{peak}-1)}{2.24j\eta_0\theta(a_s/s)^4},
\end{equation}
where $\tau_{peak}$ is the peak (or trough) optical depth value (allowing us to reduce $\cos(\eta \theta)$ to 1 [or -1] for simplicity). While this formula may not be perfectly accurate for wakes on an eccentric ringlet, it can still provide a useful rough estimate of the moons' masses. Masses are calculated using the $\tau_0$ and $\tau_{peak}$ of each occultation and their corresponding azimuthal separations, $\theta$, which are determined from the difference in the moonlet's inertial longitude and the occultation geometry longitudes of Table \ref{occ. geom.}. Table \ref{moon masses} lists the calculated moonlet masses for the two occultations and their approximate radii calculated using an assumed typical inner Uranian moon density\footnote{http://ssd.jpl.nasa.gov/?sat\_phys\_par} of $1.3$ g/cm$^3$. We find that for both the $\alpha$ and $\beta$ rings the perturbing moonlet is on the order of $10^{14}$ kg and $\sim 2-7$ km in radius, that is, less than $2\%$ of the mass and $\sim20\%$ of the radius of Cordelia. 

\section{Image analysis}

\floattable
\begin{deluxetable}{cccccc}
\tablecaption{$\alpha$ and $\beta$ ring imaging data \label{imagetable}}
\tablewidth{0pt}
\tablehead{
\colhead{Ring} & \colhead{Image Name} & \colhead{Mid-time} & \colhead{Phase angle} & \colhead{$\lambda_{range}$} & \colhead{$\lambda_s$} \\
\colhead{} & \colhead{GEOMED.IMG} & \colhead{(hr:min:s)} & \colhead{(deg)} & \colhead{(deg)} & \colhead{(deg)}
}
\startdata
$\alpha$ & C2675402 & 14:55:58 & 15.31 & 99.90 - 151.23 & 108.44 \\
 & C2675408 & 15:00:46 & 15.29 & 97.85 - 150.17 & 112.62 \\
 & C2675438 & 15:24:46 & 15.30 & 98.29 - 149.95 & 133.57 \\
 & C2675456 & 15:39:10 & 15.36 & 102.82 - 151.65 & 146.13 \\
 & C2676225 & 21:38:22 & 15.45 & 97.59 - 147.87 & 99.59 \\
 & C2676231 & 21:43:10 & 15.43 & 95.05 - 146.34 & 103.78 \\
 & C2676243 & 21:52:46 & 15.51 & 101.89 - 150.68 & 112.16 \\
 & C2676249 & 21:57:34 & 15.49 & 99.73 - 149.85 & 116.34 \\
 & C2676255 & 22:02:22 & 15.47 & 97.72 - 148.93 & 120.53 \\
 & C2676301 & 22:07:10 & 15.44 & 95.22 - 147.61 & 124.72 \\
 & C2676313 & 22:16:46 & 15.52 & 101.87 - 150.54 & 133.10 \\
 & C2676319 & 22:21:34 & 15.49 & 99.09 - 150.01 & 137.29 \\
 & C2676331 & 22:31:10 & 15.44 & 94.43 - 146.24 & 145.67 \\
 & C2678913 & 19:04:46 & 16.58 & 117.28 - 149.23 & 142.17 \\
 \hline
$\beta$ & C2675108 & 12:36:46 & 15.26 & 100.23 - 153.61 & 118.05 \\
 & C2675114 & 12:41:34 & 15.34 & 107.84 - 158.21 & 122.11 \\ 
 & C2675120 & 12:46:22 & 15.26 & 100.60 - 155.59 & 126.18 \\ 
 & C2675126 & 12:51:10 & 15.35 & 108.40 - 156.86 & 130.24 \\ 
 & C2675132 & 12:55:58 & 15.25 & 98.66 - 151.08 & 134.30 \\ 
 & C2675138 & 13:00:46 & 15.34 & 106.06 - 153.25 & 138.37 \\ 
 & C2675144 & 13:05:34 & 15.26 & 99.65 - 150.67 & 142.43 \\ 
 & C2675150 & 13:10:22 & 15.32 & 104.20 - 153.03 & 146.50 \\ 
 & C2675156 & 13:15:10 & 15.30 & 102.78 - 152.54 & 150.56 \\ 
 & C2675933 & 19:20:46 & 15.39 & 97.27 - 149.06 & 100.10 \\ 
\enddata
\tablecomments{Images used in Figure \ref{alphabetamosaics}. Image mid-times listed are in time after UTC 1986 January 21 00:00:00. The longitudinal scale is $\sim0.08^{\circ}$ pixel$^{-1}$, and the radial scale is $\sim65$ km pixel$^{-1}$. The table also includes the phase angle and range of inertial longitudes, $\lambda_{range}$, of each image. The longitude, $\lambda_s$, refers to the expected location of the moonlet in each image based on the location in Table \ref{moon locations}.}
\end{deluxetable}

\begin{figure*}
\includegraphics[width=.8\linewidth]{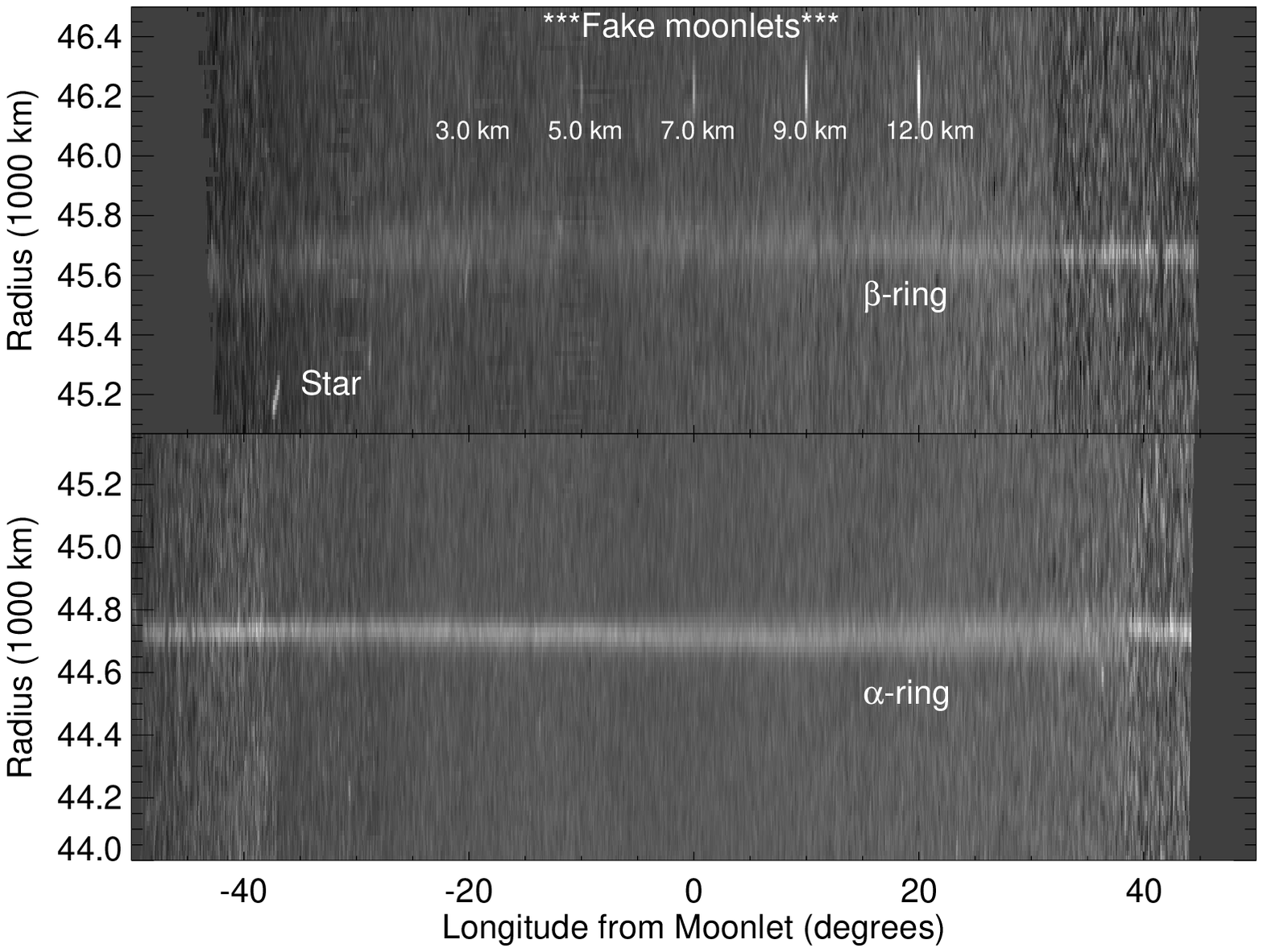}
\includegraphics[width=.8\linewidth]{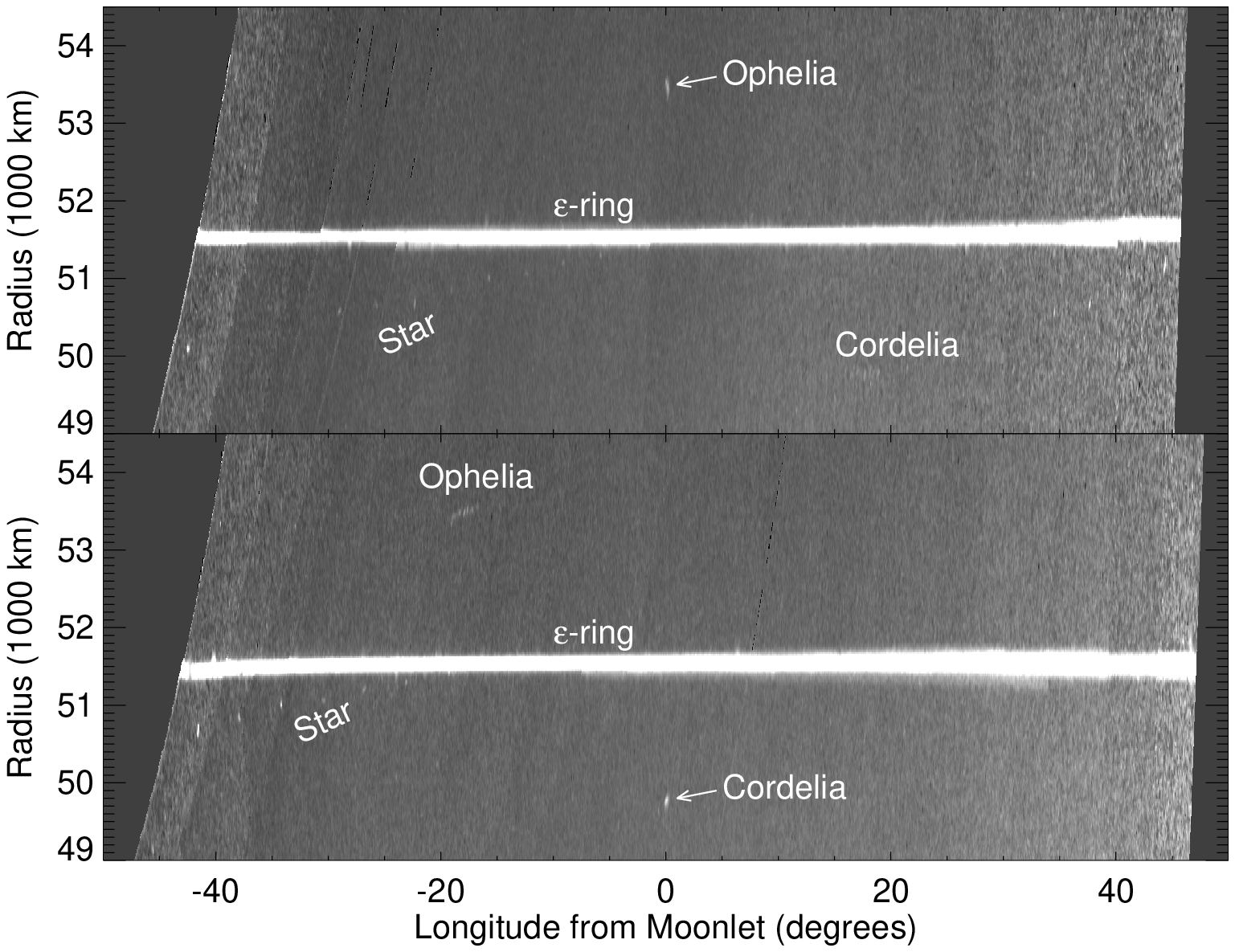}
\centering
\caption{Mosaics of the $\beta$ and $\alpha$ rings' images (top) that contain inertial longitudes within range of the predicted moonlet locations. We see no strong evidence of a moonlet just outside the $\beta$ or $\alpha$ rings. If a moonlet were present, it would be located roughly between one and two vertical scale tick marks exterior to the rings and be roughly as bright as the $3.0$ km radius fake moonlet we have inserted in the uppermost mosaic. The bottom two mosaics are a test of this technique on known moons Ophelia and Cordelia (each $\sim 20$ km in radius). Each of these mosaics also shows the other moon drifting by due to their different mean motions, as well as a background star drifting through the images. \label{alphabetamosaics}}
\end{figure*}

We can attempt to find the moonlets in the \textit{Voyager 2} images using the location estimates from the previous section. The narrow- and wide-angle camera images used for this search were obtained from the Imaging Science Subsystem on board \textit{Voyager 2}. The images are geometrically corrected and calibrated as documented on the PDS. Table \ref{imagetable} lists the images we chose for the mosaics, which contain inertial longitudes within the range of the expected moonlet locations and neglecting images containing significant defects. We re-project these images onto a corotating radius longitude grid (assuming moonlet mean motions of $1219.18 ^{\circ}/$day for $\beta$ and $1256.66 ^{\circ}/$day for $\alpha$) and co-add pixels of the appropriate images shifted such that the possible pixels of the moonlet in each image are stacked on top of one another at $0^{\circ}$ longitude, although it could be up to several tens of degrees from this location (possibly more for the $\alpha$ moonlet). These mosaics, shown in Figure \ref{alphabetamosaics}, do not show any clear evidence of a moonlet near the expected locations. We have verified that our codes do work for the known moons Cordelia and Ophelia (see bottom panel of Figure \ref{alphabetamosaics}).

Using the moonlet radii of Table \ref{moon masses}, we can determine whether the putative moonlets are too small and dim to be seen in the \textit{Voyager} images. Assuming a geometric albedo of $0.07$ \citep{2001Icar..151...51K}, we can compute the expected brightness of moons with various sizes. At the top of Figure \ref{alphabetamosaics} we show the expected signals from moons of various radii. The fake $3.0$ km moonlet is nearly indistinguishable from the many other noise features throughout the mosaic. We therefore conclude that moons with radii between 2 and 4 km are near the noise limit of the \textit{Voyager 2} images, and so it is not surprising that the moons are not easily seen in these data. We caution the reader to appreciate the uncertainties that arise when determining the point where two square root functions intersect, and so there is substantial uncertainty in the longitudes of the moons, complicating the efforts to locate the moons in images.

\section{Discussion}
Our attempts to visually detect the moonlets are not exhaustive, but given the small predicted sizes of the $\alpha$ and $\beta$ moonlets, a convincing detection may not be possible in the \textit{Voyager 2} images. Future earth-based observations may be more likely to detect these moons. Regardless of the current lack of visual detection, the identification of these periodic structures in the outer regions of the $\alpha$ and $\beta$ rings is evidence of interactions with nearby perturbers.

With their analysis of the resonances of Cordelia and Ophelia and the shepherding of the $\epsilon$ ring, \citet{1987AJ.....93..730G} theorized that a single moon, smaller than in the standard shepherding model, orbiting near a ring edge could keep the ring confined using a mechanism called angular momentum flux reversal. The theory of this mechanism, outlined in \citet{1986Icar...68..522B}, says that satellite perturbations can reverse the direction of the vicious flux of angular momentum and could possibly act over the entire width of narrow rings. \citet{2011Icar..213..201L} simulated this effect and found that collisional damping of satellite wakes caused by a small moon could keep a narrow ring confined under the right conditions. Further studies need to be done to determine whether the moonlet masses and locations found in our study are able to confine the rings.

This work was supported by the NASA Solar Systems Workings program grant NNX15AH45G. We would like to thank Mark Showalter for his valuable insights on the detection of the proposed moonlets and Richard French for several discussions concerning the \textit{Voyager 2} occultation observations and updates on the Uranian ring orbital elements. We also thank Wes Fraser for his helpful review to improve this manuscript.

\appendix
\section{Wavelength determination for RSS $\alpha$ ingress} \label{appendixA}
\begin{figure*}
\includegraphics[width=.8\linewidth]{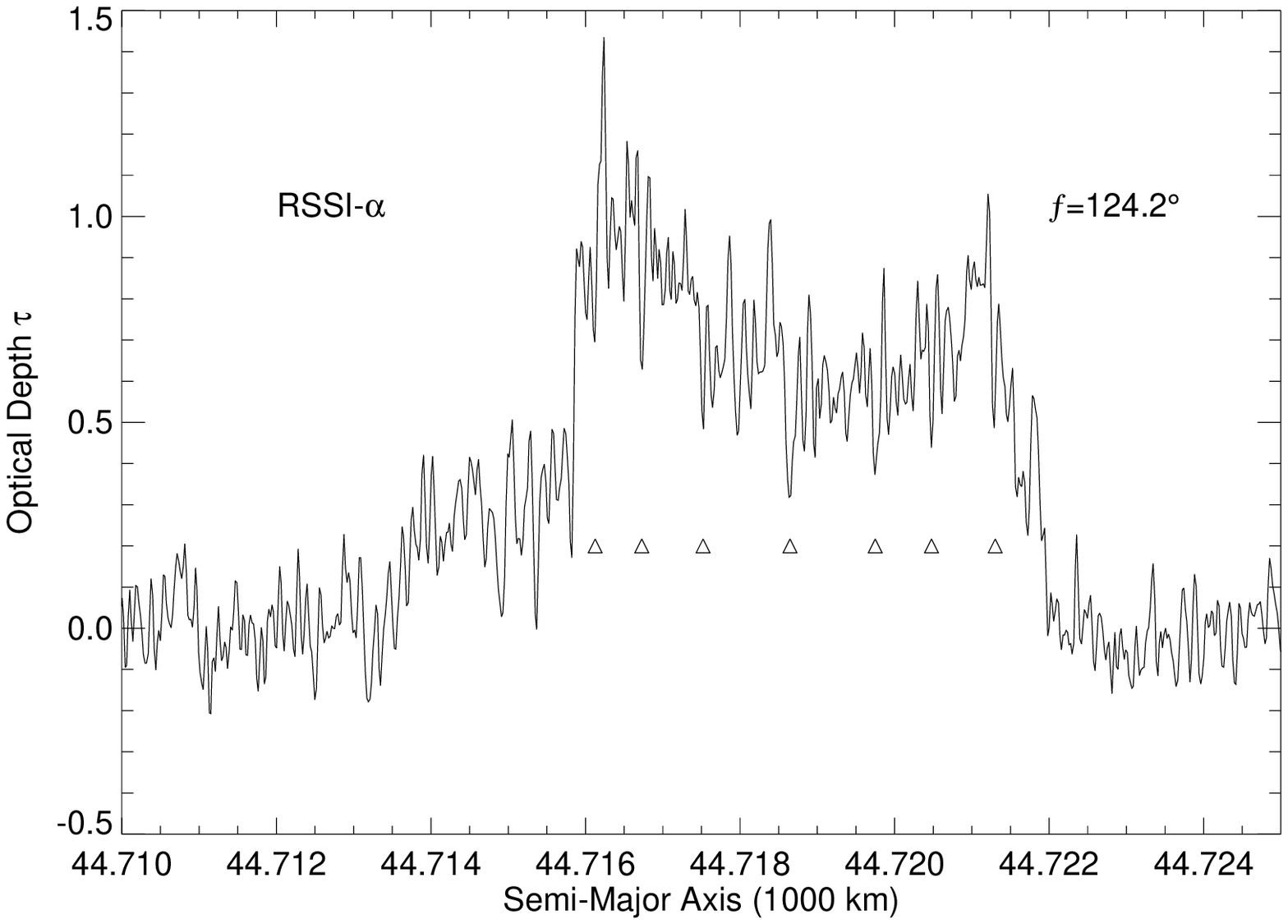}
\centering
\caption{The $\alpha$ ingress occultation in ``semimajor axis space''. We have marked the locations of the dips used in the wavelength analysis with triangles and have tabulated the exact semimajor axis of each dip in Table \ref{aimarkstable}.  \label{aimarked}}
\end{figure*}

The $\alpha$ ingress occultation scan does not contain a clear wake structure as in the other scans. As shown in Figure \ref{alpha wavelet}, this scan's wavelet transform contains no obvious locations of strong periodic structure consistent with a moonlet wake. However, the scan does contain a series of quasi-periodic dips and peaks in optical depth that are not cleanly isolated by the wavelet analysis. We determine the wake wavelength in this case by taking the average separation of these dips, marked in Figure \ref{aimarked} and listed in Table \ref{aimarkstable}. An argument could be made to use an alternate set of dips or peaks for this calculation, but we have found that the various choices result in similar-enough wavelengths. Therefore, we believe that this estimation is a reasonable approach given the limited data set. We calculate a mean wavelength of $0.86\pm0.09$ km. The uncertainty we use here is the standard error of the mean and is much larger than the uncertainty determined for the other scans. This large uncertainty extends the possible range of moonlet radius to as high as $7$ km; however, the moonlet would probably need to be on the smaller side of the range in Table \ref{moon masses} to have avoided detection in the images. 

\floattable
\begin{deluxetable}{cc}
\tablecaption{$\alpha$ ingress dip locations \label{aimarkstable}}
\tablewidth{0pt}
\tablehead{
\colhead{Semimajor Axis (km)} &
\colhead{Radius (km)}
}
\startdata
44,721.300 & 44,741.191 \\
44,720.475 & 44,740.150 \\
44,719.750 & 44,739.233 \\
44,718.650 & 44,737.842 \\
44,717.525 & 44,736.421 \\
44,716.725 & 44,735.415 \\
44,716.125 & 44,734.655 \\
\enddata
\tablecomments{Locations of dips used for wavelength analysis in both semimajor axis and radius.}
\end{deluxetable}

\section{$\beta$ Persei PPS Occultation Analysis} \label{appendixB}
\begin{figure*}
\includegraphics[width=.95\linewidth]{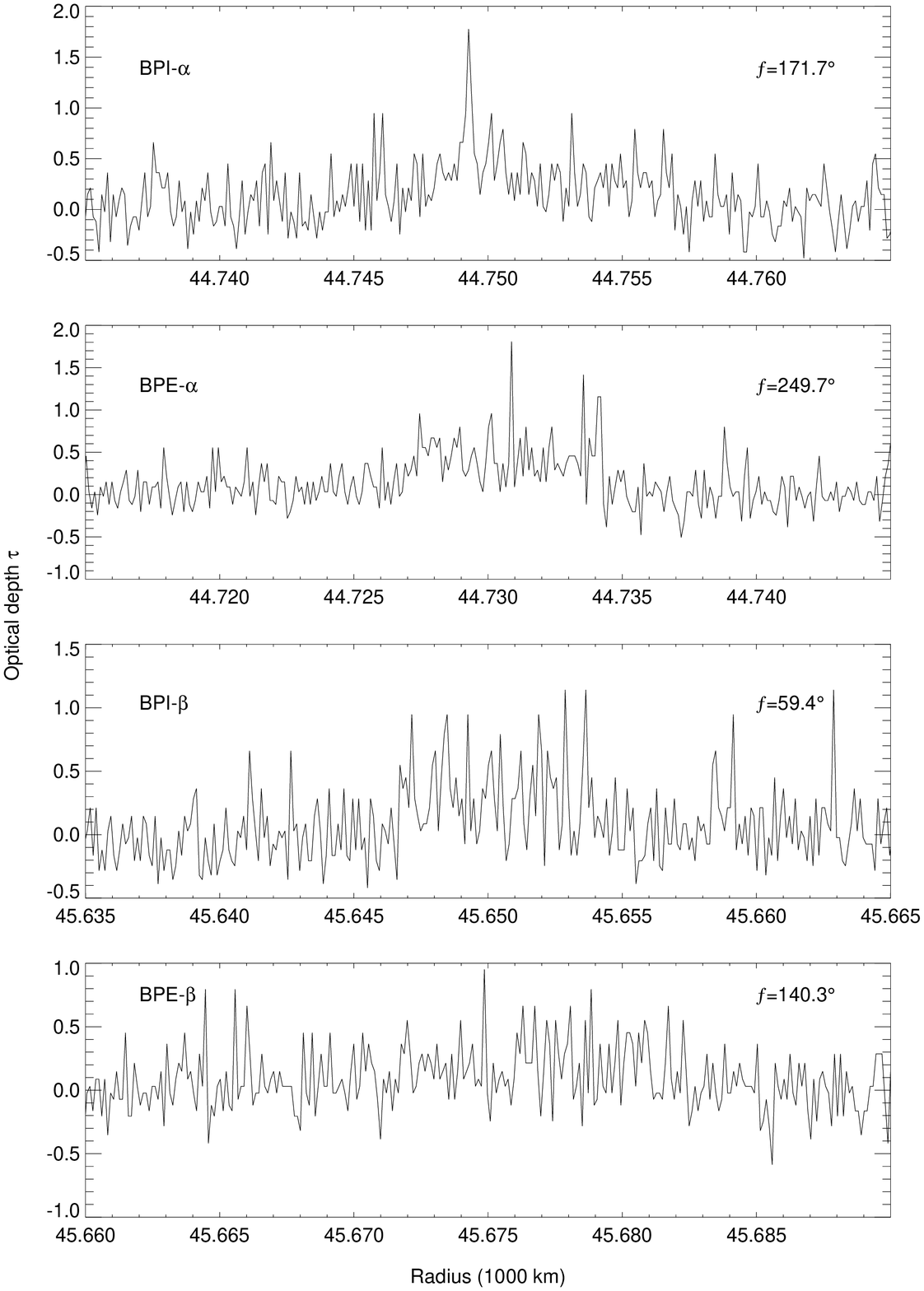}
\centering
\caption{$\beta$ Persi PPS occultations of the $\alpha$ and $\beta$ rings from the PDS. Note that like in Figure \ref{ring occ.'s}, these are the raw radial scans with varying scales in both optical depth and radius. The signal-to-noise ratio is much lower than the RSS data, so much so that it is very difficult to define edges in most cases. The $\beta$ ring is hardly detectable at all in the egress scan. \label{pps occ plot}}
\end{figure*}

\begin{figure*}
\includegraphics[width=.79\linewidth]{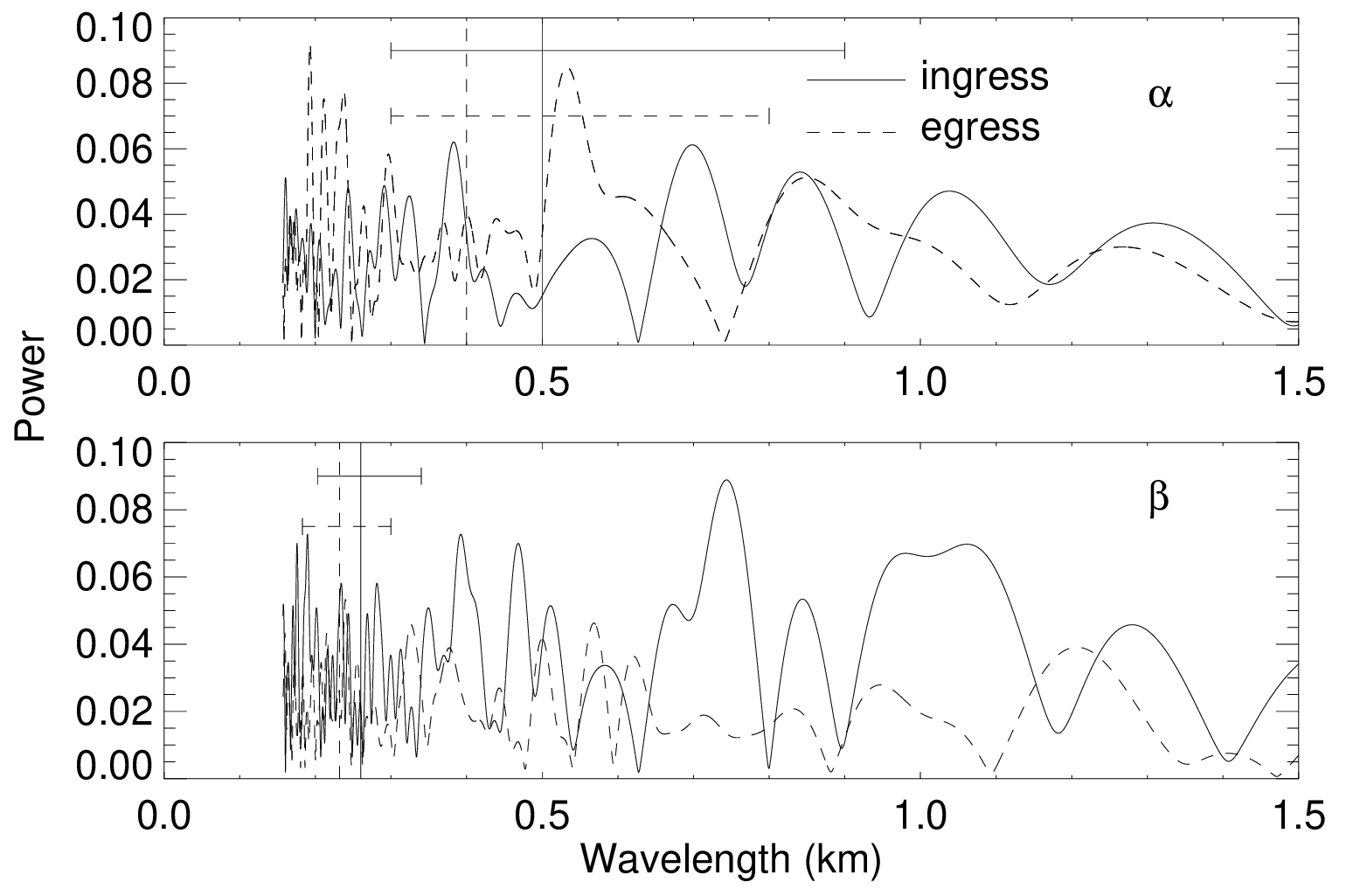}
\centering
\caption{Fourier transform of $\beta$ Persei PPS occultation scans of the $\alpha$ and $\beta$ rings. The ingress data are shown with solid lines and the egress with dashed lines. The vertical lines show the wavelengths, and uncertainties, that should be present if the moonlet is located as in the determination of the RSS analysis (see Table \ref{moon locations}). No Fourier spectra of the PPS data have dominant wavelengths (except possibly the peak at $\sim0.53$ in $\alpha$ egress, having no obvious correlation to structures in the ring profile) seen in the RSS analysis.  \label{ft pps}}
\end{figure*}

Below is a parallel analysis of the $\beta$ Persei (BP) PPS occultation observations. Table \ref{pps occ. geom.} provides a summary of the BP PPS occultation data set used here, giving mid-times, inertial longitudes and radii, and true anomalies for the $\alpha$ and $\beta$ rings at the time of the ingress and egress occultations. Figure \ref{pps occ plot} contains the unbinned raw radial occultation scans of the $\alpha$ and $\beta$ rings. The outer regions of both rings' ingress scans look to show some possible structure, but we find no dominant peaks in the Fourier spectrum of either ring. While past publications displaying binned versions of these stellar occultations \citep{1990Icar...83..102C} look to possibly contain a periodic signature, upon further inspection of the unbinned raw data we found no strong evidence of periodic optical depth variations using the methods described above for the RSS data. We show the Fourier spectra for both ingress and egress of each ring in Figure \ref{ft pps}, along with the expected wavelengths resulting from the moonlet locations determined by our RSS analysis plotted as vertical lines and listed in Table \ref{RSS wavelengths}. None of these spectra show clear, unique maxima indicative of a strong periodic signal. In fact, for both the $\beta$ ring scans the expected wavelengths are close to the occultation's sampling rate because the scan locations are significantly downstream from the last moonlet interaction. If our analysis of the RSS occultations is correct, it is unlikely that we can confirm detection of these small-wavelength optical depth variations in the $\beta$ ring stellar occultations. We also note that in the egress scan the $\beta$ ring is almost undetectable. 

\begin{figure*}
\includegraphics[width=.79\linewidth]{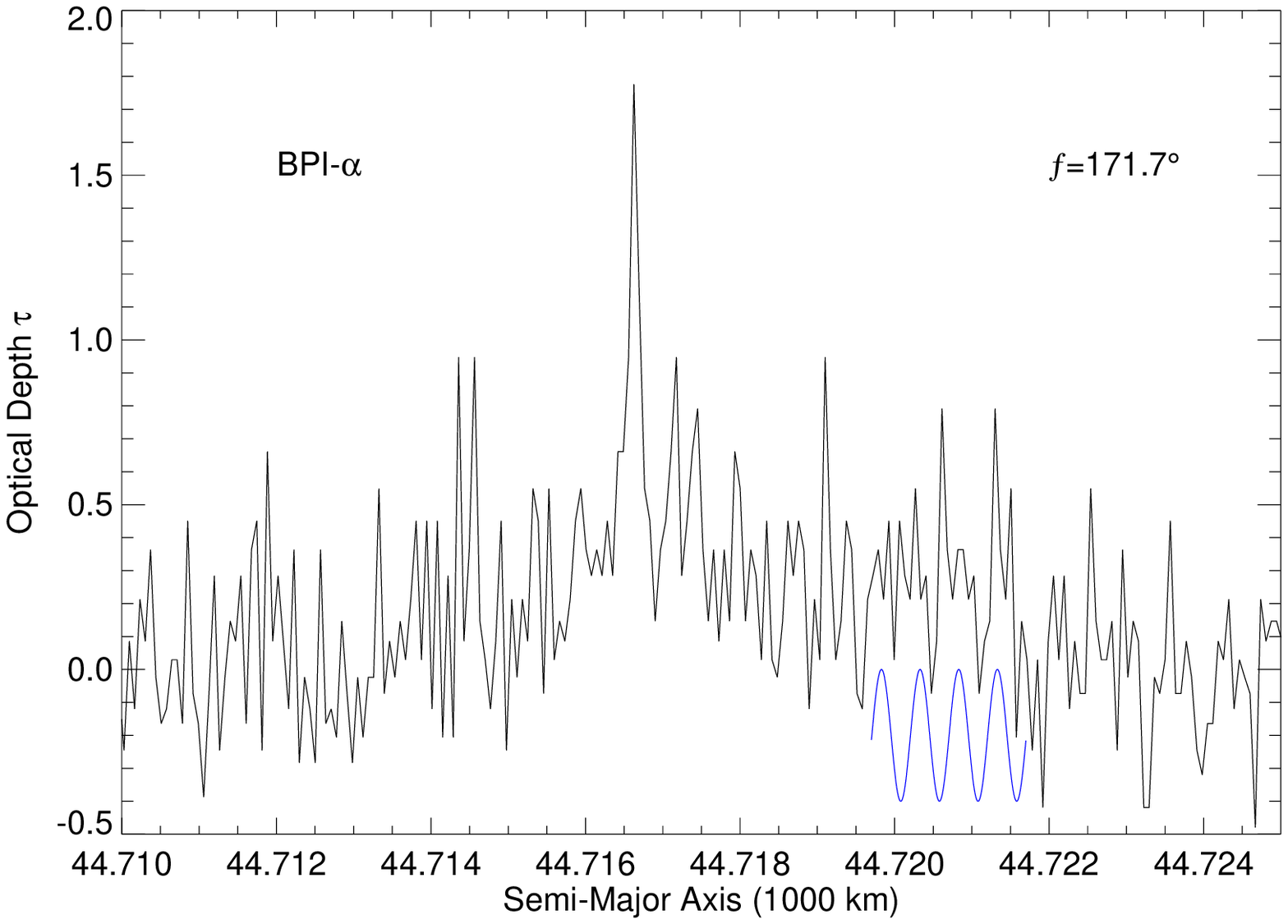}
\centering
\caption{$\beta$ Persei PPS $\alpha$ ingress scan in semimajor axis space. The blue sine wave, with wavelength of 0.5 km, plotted below the data matches reasonably well with the quasi-periodic pattern near the outer edge of the ring. Given the low signal-to-noise ratio of the PPS data set, this is the best evidence we have for a connection between these observations, a proposed perturbing moonlet, and the RSS observations. \label{alpha ingress pps}}
\end{figure*}

In the case of the $\alpha$ ring, the expected wavelengths are longer, and there are some weak peaks in the vicinity of the predictions based on the RSS data. However, these peaks are not far above the noise level and so cannot be regarded as strong evidence for periodic signals. However, we do observe a structure near the outer edge of the $\alpha$ ingress scan that arguably looks similar to the that of the RSS $\alpha$ ingress scan. Figure \ref{alpha ingress pps} shows the $\alpha$ ingress scan in semimajor axis space accompanied by a sine wave with a wavelength we would expect the wake to have here, assuming a moonlet located as in Table \ref{moon locations}. This wave appears to fit reasonably well here, but we do not claim this to be evidence that our $\alpha$ moonlet location is correct. Overall, we find that the analysis of the $\beta$ Persei PPS occultation scans cannot confirm or deny the moonlet locations of the RSS analysis.
\floattable
\begin{deluxetable}{cccccc}
\tablecaption{Geometry of $\beta$ Persei Photopolarimeter Subsystem (PPS) occultations\label{pps occ. geom.}}
\tablewidth{0pt}
\tablehead{
\colhead{Ring} & 
\colhead{Occ.} & 
\colhead{True Anom.} & 
\colhead{Mid-time} & 
\colhead{Mid-rad.} & 
\colhead{Mid-long.} \\
\colhead{} & 
\colhead{} & 
\colhead{(deg)} & 
\colhead{(hr:min:s)} & 
\colhead{(km)} & 
\colhead{(deg)} 
}
\startdata
$\alpha$ & BPI & 171.7 & 18:38:44.398 & 44,752.86 & 29.3 \\
& BPE & 249.7 & 19:33:16.498 & 44,731.10 & 107.4 \\
$\beta$ & BPI & 59.7 & 18:37:21.368 & 45,651.83 & 27.9 \\
& BPE & 140.5 & 19:34:42.938 & 45,677.36 & 108.8 \\
\enddata
\tablecomments{The appended labels of I and E stand for ingress and egress after BP for $\beta$ Persei. The true anomaly of each ring at the time of their respective ring intercept mid-times and mid-radii is calculated from the inertial longitudes provided and the updated precession rates provided by R. G. French (see Table \ref{orb elements}). Mid-times listed are the times of ring intercept measured in seconds after UTC 1986 January 24 00:00:00, when the light measured by PPS from $\beta$ Persei (Algol) intercepted the mid-ring radius. The radial position of the rings shown in Figure \ref{pps occ plot}, created using the data provided on the PDS rings node, differs slightly from older publications.} 
\end{deluxetable}

\floattable
\begin{deluxetable}{cc}
\tablecaption{Expected wavelengths for the PPS ring profiles based on the RSS analysis moonlet location. \label{RSS wavelengths}}
\tablewidth{0pt}
\tablehead{
\colhead{Ring Scan} &
\colhead{Wavelength (km)}
}
\startdata
BPI-$\alpha$ & $0.5^{+0.4}_{-0.2}$ \\
BPE-$\alpha$ & $0.4^{+0.4}_{-0.1}$ \\
BPI-$\beta$ & $0.26^{+0.08}_{-0.06}$ \\
BPE-$\beta$ & $0.23^{+0.07}_{-0.05}$ \\
\enddata
\end{deluxetable}

\bibliography{uranuspaper2}
\bibliographystyle{aasjournal}

\end{document}